\documentclass[aps,prl,twocolumn,groupedaddress,floats,showpacs]{revtex4-1}
\usepackage{latexsym}
\usepackage{dcolumn}
\usepackage[dvips]{graphicx}
\usepackage{amssymb}
\usepackage{graphics}
\usepackage{amsmath}
\usepackage{epsf}

\newcommand{\beq}    {\begin{equation}}
\newcommand{\enq}    {\end{equation}}

\begin{document}
\title{Conductivity of graphene on boron nitride substrates}
\author{S. Das Sarma and E. H. Hwang} 
\affiliation{Condensed Matter Theory Center, Department of 
        Physics, University of Maryland, College Park, MD 20742-4111}

\begin{abstract}
We calculate theoretically the disorder-limited conductivity
of monolayer and bilayer graphene on hexagonal boron nitride (h-BN) substrates, comparing
our theoretical results with the recent experimental results. The
comparison leads to a direct quantitative estimate of the underlying
disorder strength for both short-range and long-range disorder in the
graphene on h-BN system.
We find that the good interface quality between
graphene and h-BN leads to strongly
suppressed charged impurity scattering compared with the corresponding
SiO$_2$ substrate case, thus producing very high mobility for the 
graphene on h-BN system. 
%As a consequence, bilayer graphene on h-BN substrates 
%has much less disorder than the corresponding monolayer
%system in contrast to graphene on SiO$_2$ substrates where the bilayer
%system is typically more strongly disordered. 
\end{abstract}
\maketitle

%\section{introduction}

An important recent development in the physics and materials science
of graphene \cite{dassarma2010,review09} is the successful fabrication
of gated graphene layers on hexagonal 
boron nitride (h-BN) substrates\cite{dean2010a,dean2010b}. Since h-BN has
the same hexagonal 
honeycomb lattice structure as graphene itself with an almost matching
lattice constant, the expectation has been that graphene on h-BN would
have much lower disorder than the generic graphene on SiO$_2$
substrates that has almost universally been studied so far
experimentally. This expectation has, in fact, been spectacularly
borne out by the recent experiments \cite{dean2010a,dean2010b} at
Columbia University where both 
monolayer graphene (MLG) and bilayer graphene (BLG) on h-BN substrates
have been shown to
have substantially higher (by roughly one order of magnitude or more)
carrier mobility than graphene samples on the standard SiO$_2$
substrates \cite{dassarma2010,review09}. In fact, the quality of
graphene on h-BN, as measured by 
transport experiments, appears to be comparable to that of annealed
suspended graphene\cite{ghahari2010,bao2010}, with both systems
exhibiting clear fractional 
quantum Hall effects attesting to their very high mobility.

In the current work, we consider theoretically electronic transport
properties of graphene/BN MLG and BLG systems, using the highly
successful Boltzmann-Kubo-RPA formalism which has earlier been
used\cite{dassarma2010} to
study graphene transport on SiO$_2$ substrates, both for MLG
\cite{hwang2007,adam2007,hwang2009} and BLG \cite{dassarma2010b}
systems, as well as for the suspended graphene \cite{adam2008}
system. Our goal is a 
thorough quantitative understanding of the specific operational features
of resistive scattering mechanisms limiting carrier mobility in
graphene on h-BN. By demanding quantitative agreement between our calculated
graphene (on h-BN) transport properties with the corresponding
experimental data \cite{dean2010a,dean2010b} for both MLG and BLG
systems, we establish the 
precise role of long-range (e.g. charged Coulomb impurities in the
environment) versus short-range (e.g. point defects, neutral
scatterers, vacancies) disorder in graphene on h-BN systems. We obtain
excellent agreement with the experimental data
\cite{dean2010a,dean2010b} using very reasonable
disorder parameters, establishing that the good interface quality between
graphene and h-BN (e.g. lack of dangling bonds) leads to strongly
suppressed charged impurity scattering compared with the corresponding
SiO$_2$ substrate situation \cite{hwang2007,adam2007}, thus providing
very high mobility for the 
graphene on h-BN system. The relative suppression of long-range
scattering compared with the short-range scattering also leads to
rather nonlinear-looking MLG conductivity as a function of gate
voltage (i.e. carrier density) for graphene on h-BN substrates compared
with the SiO$_2$ substrates, thus explaining the peculiar experimental
finding that the observed BLG (MLG) conductivity on h-BN substrates
\cite{dean2010a,dean2010b} 
manifests linear (nonlinear) conductivity as a function of the gate
voltage. Our theory also naturally explains the weaker observed
temperature dependence of MLG conductivity than the BLG conductivity
for graphene on h-BN substrate. 
The actual conductivity of MLG/BN or BLG/BN \cite{dean2010a,dean2010b}
is determined by the detailed interplay between the long-range and the
short-range disorder in the relevant system along with the distinct
screening properties of the graphene carriers as it is in the usual
graphene on SiO$_2$ substrates
\cite{dassarma2010,hwang2007,adam2007,dassarma2010b,xiao2010}. 

%One important implication of the direct
%qualitative comparison between our theory and the experimental data
%for graphene on h-BN substrates is that the effective disorder appears
%to be much lower in the BLG samples of the existing Columbia samples
%\cite{dean2010a,dean2010b} 
%than in the corresponding MLG samples, which is unexpected in view of
%the fact that exactly the reverse seems to be the case for graphene on
%SiO$_2$ substrates \cite{dassarma2010b,xiao2010} where typically BLG
%systems have much higher 
%(lower) disorder (mobility) than the MLG systems.

The graphene conductivity $\sigma$ is given by \cite{dassarma2010}
\begin{equation}
\sigma = \frac{e^2}{2}\int d\varepsilon D(\varepsilon)v_k^2
\tau(\varepsilon) \left (- \frac{\partial f}{\partial \varepsilon}
\right ),
\end{equation}
where $f=f(\varepsilon_k)$ is the Fermi distribution function,
$D(\varepsilon)$ is the density of states, $v_k=d\varepsilon_k/dk$ is
the carrier velocity, and $\tau(\varepsilon)$ is the transport
scattering (or relaxation) time which depends explicitly on the
effective disorder scattering potential `$V$':
\begin{widetext}
\begin{equation}
\frac{1}{\tau(\varepsilon)} = \frac{2\pi}{\hbar} \sum_{\alpha}
n_i^{(\alpha)}(z) \int \frac{d^2k'}{(2\pi)^2} \left | V_{{\bf kk}'}(z)
\right |^2 g(\theta_{{\bf kk}'})(1-\cos \theta_{{\bf kk}'}) \delta
(\varepsilon_{\bf k}-\varepsilon_{{\bf k}'}),
\end{equation}
\end{widetext}
where $\varepsilon_{\bf k}$ is the graphene carrier energy dispersion for 2D
wave vector {\bf k}, `z' is the position of the impurity whose
concentration is defined by $n_i^{\alpha}$ with $\alpha$ denoting the
kind of impurity (e.g. long-range or short-range), $g(\theta)$ denotes a
known chiral matrix element form factor determined by the band
structure (and is therefore different for MLG and BLG), and ({\bf k},
${\bf k}'$) are the incoming and outgoing carrier 2D wave vector due to
the impurity scattering potential $V_{{\bf kk}'}(z)$. Since the details of the transport
theory for graphene have been discussed earlier in the literature
\cite{dassarma2010,hwang2007,hwang2009,dassarma2010b}, we
only make a few comments on the calculational aspects of our
theoretical results presented in this work: (i) the substrate h-BN is
characterized by its static dielectric constant $\kappa_{BN}=7.0$ \cite{geick1966},
leading to an effective background dielectric constant $\kappa=4.0$,
which enters into the definition of the effective disorder potential;
(ii) the effective disorder potential `$V$' entering Eq. (2) is taken to
be the screened disorder where the screening is by the static graphene
(MLG or BLG) dielectric function $\epsilon(q,T)$ which has been
calculated earlier in refs.~\cite{hwang2007b,hwang2008b}  respectively
for MLG and BLG; (iii) we 
include two types of disorder in our theory, the long-range disorder
characterized by randomly distributed charged impurity centers with
2D density of $n_i$ located at the graphene-BN interface and the
short-range disorder characterized by an effective strength of
$n_dV_0^2$ denoting a white-noise delta-correlated local disorder. (We
emphasize that both long- and short-range disorder are necessary for
quantitative and
qualitative understanding of experimental data.)

%%%%%%%%%%%%%%%%%%%%%Fig.1 %%%%%%%%%%%%%%%%%%%%%%%
\begin{figure}
\includegraphics[width=0.8\columnwidth]{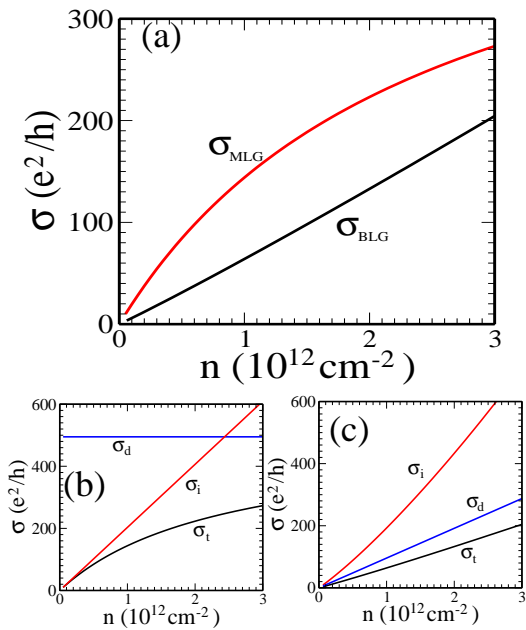}
\caption{ (Color online)
(a) Calculated conductivity of MLG ($\sigma_{MLG}$) and BLG
  ($\sigma_{BLG}$) using the following parameters: $n_i=12\times
  10^{10}cm^{-2}$ and $n_dV_0^2=0.7$ (eV\AA)$^2$.
In (b) [(c)] we show the individual MLG [BLG] conductivity limited by
long-range scattering ($\sigma_i$) 
and short-range scattering ($\sigma_d$). $\sigma_t$ indicates the
total conductivity limited by the both scatterings. 
\label{fig:1}
}
\end{figure}
%%%%%%%%%%%%%%%%%%%%%%%%%%%%%%%%%%%%%%%%%%%%%

The theory is characterized by two parameters, $n_i$ and $n_dV_0^2$,
describing long-range and short-range disorder, respectively. In
principle, the effective separation ($d$) between the location of the
charged impurity centers and the 2D graphene layer could also be an
additional physically relevant parameter in the transport theory
\cite{dassarma2010,hwang2007,adam2007}, but
we put $d=0$ throughout this paper, keeping the number of free
parameters a minimum (only two) and assuming that the random charged
impurity centers are located at the graphene/BN interface as
consistent with the very high quality of the h-BN crystals used in
refs. \cite{dean2010a,dean2010b}. For obtaining our theoretical
transport results we have varied 
the parameters $n_i$ and $n_dV_0^2$ arbitrarily over a wide range,
obtaining the best regression fit to the high-density data of
refs.~\cite{dean2010a,dean2010b} as shown in Figs.~1 and 2.
(We have also used different values of '$d$' using the charged impurity
separation as a tuning parameters, but our results are qualitatively
unaffected by an adjustable $d$.)

We first show our theoretical results valid at ``high'' carrier
density ($n$) defined as $n \agt n_i$ away from the minimum
conductivity Dirac point regime, where the density fluctuations
associated with the inhomogeneous puddle formation can be safely
neglected. In Figs.~1 and 2 we show our calculated conductivity (at
$T=0$), $\sigma(n)$, as a function of the carrier density for a few
different values of the disorder parameters choosing the parameters
such that we get essentially exact quantitative agreement away from
the Dirac point ($n > n_i$) with the experimental data of
ref.~\cite{dean2010a} for 
MLG/BN (Fig.1) and BLG/BN (Fig.2) systems. In each figure, we present
results for both MLG and BLG systems for a fixed set of values of the
disorder parameters $n_i$ (long-range) and $n_dV_0^2$ (short-range)
with the results of Fig. 1 (2) showing quantitative agreement with the
corresponding experimental data for MLG (BLG) on h-BN in
ref.~\cite{dean2010a}. In each
figure we present the individual conductivity limited by long-range
and short-range scattering as well as the total conductivity.

%%%%%%%%%%%%%%%%%%%%%Fig.2 %%%%%%%%%%%%%%%%%%%%%%%
\begin{figure}
\includegraphics[width=0.8\columnwidth]{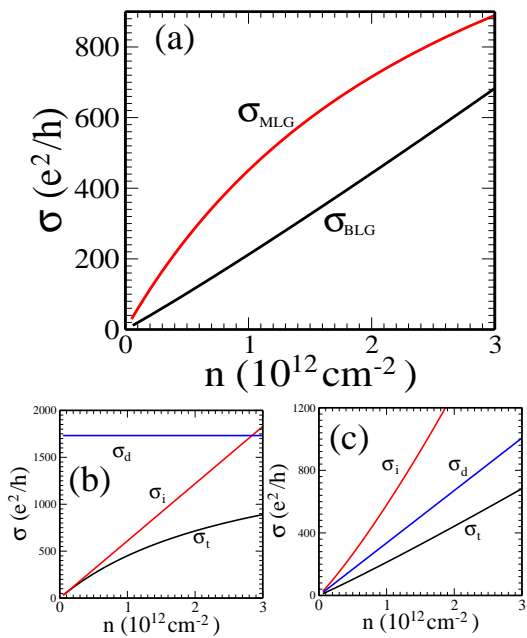}
\caption{ (Color online)
(a) Calculated conductivity of MLG ($\sigma_{MLG}$) and BLG
  ($\sigma_{BLG}$) using the following parameters: $n_i=4\times
  10^{10}cm^{-2}$ and $n_dV_0^2=0.2$ (eV\AA)$^2$.
In (b) [(c)] we show the individual MLG [BLG] conductivity limited by
long-range scattering ($\sigma_i$) 
and short-range scattering ($\sigma_d$). $\sigma_t$ indicates the
total conductivity limited by the both scatterings. 
\label{fig:2}
}
\end{figure}
%%%%%%%%%%%%%%%%%%%%%%%%%%%%%%%%%%%%%%%%%%%%%

Three qualitative features of our theoretical results in Figs.~1 and 2
stand out: (i) for fixed disorder, MLG conductivity is always larger
than BLG conductivity for all densities although they approach each
other at very high density as expected; (ii) the quantitative
values of the disorder parameters (i.e. $n_i$ and $n_dV_0^2$)
necessary in Figs.~1 and 2 for obtaining agreement with the
experimental data \cite{dean2010a,dean2010b} for graphene on h-BN
substrates are typically much (by 
more than an order of magnitude) smaller than that needed for
agreement between theory and experiment with the corresponding
graphene on SiO$_2$ substrates (e.g. refs.~\cite{hwang2007,adam2007})
--- this is particularly 
true for the charged impurity density $n_i$ which has the remarkably small
value of $0.3 \times 10^{11} - 1.0 \times 10^{11}$ cm$^{-2}$ for
graphene on h-BN substrates compared with $n_i > 10^{12}$ cm$^{-2}$ for
graphene on SiO$_2$ substrates (we note that short-range disorder
characterized by $n_dV_0^2$ seems comparable in strength for h-BN and
SiO$_2$ substrates with h-BN having somewhat smaller values); (iii) the
MLG conductivity results for h-BN substrates are much more sublinear than for the
corresponding SiO$_2$ substrate case clearly establishing the much
weaker role of long-range charged impurity scattering in h-BN systems
compared with SiO$_2$ systems.

It is easy to show theoretically
\cite{dassarma2010,hwang2007,adam2007} using Eq.~(2) that the charged
impurity scattering limited MLG conductivity
$\sigma_i^{MLG}$ on h-BN substrates is given approximately by
$\sigma_i^{MLG} \approx 25.7 (e^2/h) (n/n_i)$ whereas the short-range
scattering limited conductivity is given by $\sigma_d^{MLG} \approx
350 (e^2/h)/(n_dV_0^2)$ where $n_dV_0^2$ is measured in (eV\AA)$^2$
units. Our MLG numerical results for long-range scattering shown in
Figs.~1 and 2 obey these analytical relations exactly with the net
conductivity being given by $\sigma =
(\sigma_i^{-1}+\sigma_d^{-1})^{-1}$. For the BLG on h-BN substrates, a
simple analytic relation can only be derived for the short-range
scattering-limited conductivity $\sigma_0^{BLG} = 66.7
(e^2/h)(n/n_dV_0^2)$, which is linear in carrier density, with '$n$' in
units of $10^{12}$ cm$^{-2}$ and $n_dV_0^2$ in units of (eV\AA)$^2$. The
long-range disorder leads to a $\sigma_i^{BLG} \sim n^{\alpha}$ where
$\alpha \approx 1-1.3$ depending on the parameter regime, and no
simple analytic relationship can be derived except at very low BLG
carrier density where the Coulomb disorder is effectively completely
screened out since the BLG screening wave vector becomes much larger
than the Fermi wave vector. For this very low carrier density regime
($\ll 10^{12}$ cm$^{-2}$), the charged impurity disorder limited BLG
conductivity becomes linear in carrier density (i.e., $\alpha=1$)
obeying the approximate relationship: $\sigma^{BLG}_i \sim 15 (e^2/h)
(n/n_i)$. We mention, however, that this formula is not useful for $n
< n_i$ since density inhomogeneity effects associated with puddles
would dominate close to the charge neutrality point.

The agreement, using reasonable values of disorder parameters, between
theoretical results presented in Figs.~1 and 2 
with the experimental data \cite{dean2010a,dean2010b} of the Columbia
group indicates that 
graphene on h-BN indeed has substantially lower long-range Coulomb
disorder in its environment than graphene on SiO$_2$ substrates, most
likely due to the high-quality graphene/BN interface without any
dangling bonds as already speculated in ref.~\cite{dean2010a}. A
direct consequence of 
this reduced Coulomb scattering is the manifestly sublinear
$\sigma(n)$ observed in the MLG/BN system to be contrasted with the
linear $\sigma(n)$ in the MLG/SiO$_2$ system
\cite{dassarma2010,hwang2007} except at very high
densities. We note that our theory indicates a direct way of
estimating the strength of both long- and short-range disorder from
the high-density MLG/BN $\sigma(n)$ data by obtaining the slope
$d\sigma/dn$ at high-density (which gives $n_i$) and by obtaining the
intercept of the high-density $\sigma(n)$ extrapolated to
$n \rightarrow 0$, which gives $n_dV_0^2$.

%%%%%%%%%%%%%%%%%%%%%Fig.3 %%%%%%%%%%%%%%%%%%%%%%%
\begin{figure}
\includegraphics[width=\columnwidth]{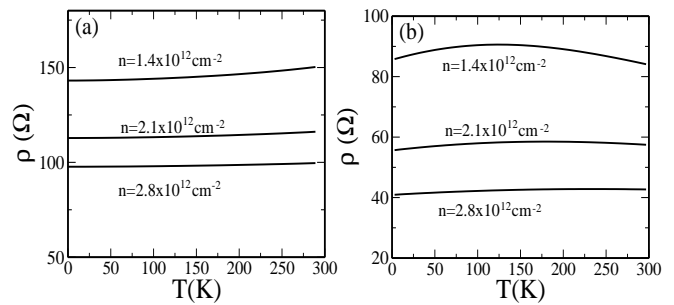}
\caption{
The temperature dependence of the MLG conductivity (a) and the BLG
conductivity (b) for several electron densities. In (a) [(b)] the parameters
of Fig.~1 [Fig.~2] are used.
\label{fig:3}
}
\end{figure}
%%%%%%%%%%%%%%%%%%%%%%%%%%%%%%%%%%%%%%%%%%%%%

In Fig.~3, we present our theoretical results for the
temperature dependence of the MLG (Fig.~3) and BLG (Fig.~4)
conductivity, $\sigma(n,T)$, on h-BN substrates. These results are again
valid (similar to those in Figs.~1 and 2) at high carrier density ($n
> n_i$) where density inhomogeneity effects are negligible. All
phonon effects \cite{hwang2008c,efetov2010} are neglected here with
the temperature dependence 
arising entirely from the temperature dependence of the screening
function and the energy-averaging associated with the
finite-temperature smearing of the Fermi surface
\cite{hwang2009}. The first effect 
(``screening'') produces weak metallic temperature dependence
(i.e. $\sigma$ decreasing with increasing $T$) since screening weakens
at higher temperatures whereas the second effect (``thermal
averaging'')  produces weak insulating temperature dependence
(i.e. $\sigma$ decreasing with increasing $T$). Although the
temperature dependence is weak, as is obvious from Fig.~3,
the theoretical behavior of $\sigma(T)$ is qualitatively consistent
with the experimental observations \cite{dean2010a} away from the
Dirac point: (i) 
MLG manifests weak metallic T-dependence, and (ii) BLG manifests weak
insulating T-dependence. Experimentally, both systems manifest
insulating $\sigma(T)$ at the Dirac point where density
inhomogeneity effects dominate \cite{hwang2010}, but at higher
density ($n \agt n_i$) 
our results are consistent with experimental finding of
ref.~\cite{dean2010a}.

%%%%%%%%%%%%%%%%%%%%%Fig.4 %%%%%%%%%%%%%%%%%%%%%%%
\begin{figure}
\includegraphics[width=\columnwidth]{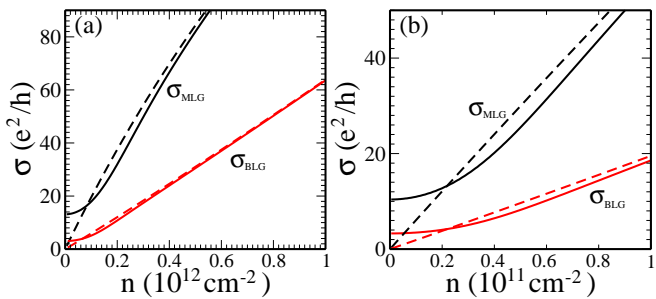}
\caption{ (Color online)
Calculated conductivity using effective medium theory (solid lines)
for both MLG ($\sigma_{MLG}$) and ($\sigma_{BLG}$).
Dashed lines represent the transport results calculated using
Boltzmann transport theory.
In (a) and (b) we use the parameters corresponding to the Fig.~1 and
2, respectively. Note that the effective medium theory becomes
important only at low ($n<n_i$) carrier density where inhomogeneous
puddle formation becomes significant.
\label{fig:4}
}
\end{figure}
%%%%%%%%%%%%%%%%%%%%%%%%%%%%%%%%%%%%%%%%%%%%%

Finally, we consider in Fig.~4 the low-density transport in graphene/BN systems,
where the results (valid for $n \agt n_i$) shown in Figs.~1--3 do not
apply. At low carrier density $n$ ($<n_i$), which is very low ($\sim
10^{10}cm^{-2}$) for the graphene/BN system because of its extremely
weak Coulomb disorder, the graphene layer is known
\cite{dassarma2010,adam2007,rossi2008,martin2008,zhang2009,deshpande2009} to break up into
inhomogeneous puddles due to the failure of screening, and thus the
naive Boltzmann-Kubo-RPA transport theory, which explicitly assumes a
homogeneous carrier density, is no longer valid since the density
fluctuations become strong. To demonstrate the effect of puddle
formation on graphene/BN transport properties, we have carrier out an
effective medium theory \cite{dassarma2010b,rossi2009} calculation of
transport using $n_{rms}=n_i$, 
where $n_{rms}$ is the root-mean-square fluctuation in the carrier
density due to the puddles induced by the charged
impurities. Microscopic self-consistent calculations
\cite{dassarma2010b,rossi2008,rossi2009} show that 
$n_{rms}\approx n_i$ is a reasonable qualitative approximation for the
density inhomogeneity around the Dirac point. 
Our $T=0$ effective
medium theory transport results (using the Boltzmann-Kubo-RPA
transport formalism) are shown in Fig.~5 for both MLG/BN and BLG/BN
systems. The most important features of Fig.~5 are: (i) the high
density ($n \agt n_i$) results shown in Figs.~1 and 2 remain valid;
(ii) near the Dirac point (for $n<n_i$), $\sigma(n)$ saturates with a
non-universal disorder-dependent minimum conductivity
\cite{dassarma2010,adam2007,rossi2009} whose value is
roughly given by $2-10$ ($e^2/h$) consistent with the experimental
observations \cite{dean2010a,dean2010b}.

We have also carried out our effective medium theory calculation at
finite temperature to include the puddle effects on
$\sigma(n,T)$. These results (not shown) agree with the results shown
in Figs.~3 and 4 at high densities ($n>n_i$), but for low carrier
densities ($n<n_i$) we qualitatively recover the experimentally
observed strongly insulating $\sigma(T)$ induced by the puddles.

This work is funded by US-ONR.

%-------------------------------------------------------

%\bibliographystyle{apsrev}
%\bibliography{bib}
%\bibliography{journal-abbreviations,bib}

\end{document}